\documentclass[twocolumn]{aastex631}

\newcommand{\mathbold}[1]{\mbox{\boldmath $\bf#1$}}
\newcommand\piEbold{{\mathbold \pi_{\rm E}}}
\newcommand\mubold{{\mathbold \mu}}
\newcommand\Vbold{{\mathbold V_{\oplus,\perp}}}

\usepackage{verbatim}
\usepackage{amsmath}
\received{??}
\revised{??}
\accepted{??}

\shorttitle{OB13-132Lb: Saturn-Mass Planet Around an M Dwarf}
\shortauthors{Rektsini et al.}

\graphicspath{{./}}

\begin{document}

\title{Precise mass measurement of OGLE-2013-BLG-0132/MOA-2013-BLG-148: a Saturn mass planet orbiting an M-dwarf }

\author[0000-0002-1530-4870]{Natalia E. Rektsini}
\affiliation{School of Natural Sciences,
University of Tasmania,
Private Bag 37 Hobart, Tasmania, 7001, Australia}

\affiliation{Sorbonne Universit\'e, CNRS, Institut d'Astrophysique de Paris, IAP, F-75014, Paris, France}

\author{Virginie Batista}
\affiliation{Sorbonne Universit\'e, CNRS, Institut d'Astrophysique de Paris, IAP, F-75014, Paris, France}

\author[0000-0003-2388-4534]{Clément Ranc}
\affiliation{Sorbonne Universit\'e, CNRS, Institut d'Astrophysique de Paris, IAP, F-75014, Paris, France}

\author{David P. Bennett}
\affiliation{Code 667, NASA Goddard Space Flight Center, Greenbelt, MD 20771, USA}
\affiliation{Department of Astronomy, University of Maryland, College Park, MD 20742, USA}

\author[0000-0003-0014-3354]{Jean-Philippe Beaulieu}
\affiliation{School of Natural Sciences,
University of Tasmania,
Private Bag 37 Hobart, Tasmania, 7001, Australia}
\affiliation{Sorbonne Universit\'e, CNRS, Institut d'Astrophysique de Paris, IAP, F-75014, Paris, France}

\author[0000-0001-5860-1157]{Joshua W. Blackman}
\affiliation{School of Natural Sciences,
University of Tasmania,
Private Bag 37 Hobart, Tasmania, 7001, Australia}
\affiliation{Physikalisches Institut, University of Bern, Gesellschaftsstrasse 6, 3012 Bern, Switzerland}

\author[0000-0003-0303-3855]{Andrew A. Cole}
\affiliation{School of Natural Sciences,
University of Tasmania,
Private Bag 37 Hobart, Tasmania, 7001, Australia}

\author{Sean K. Terry}
\affiliation{Department of Astronomy, University of California Berkeley, Berkeley, CA 94701, USA}
\affiliation{Code 667, NASA Goddard Space Flight Center, Greenbelt, MD 20771, USA}
\affiliation{Department of Astronomy, University of Maryland, College Park, MD 20742, USA}

\author{Naoki Koshimoto}
\affiliation{Department of Earth and Space Science, Graduate School of Science, Osaka University, Toyonaka, Osaka 560-0043, Japan}
\affiliation{Code 667, NASA Goddard Space Flight Center, Greenbelt, MD 20771, USA}
\affiliation{Department of Astronomy, University of Maryland, College Park, MD 20742, USA}

\author{Aparna Bhattacharya}
\affiliation{Code 667, NASA Goddard Space Flight Center, Greenbelt, MD 20771, USA}
\affiliation{Department of Astronomy, University of Maryland, College Park, MD 20742, USA}

\author{Aikaterini Vandorou}
\affiliation{Code 667, NASA Goddard Space Flight Center, Greenbelt, MD 20771, USA}
\affiliation{Department of Astronomy, University of Maryland, College Park, MD 20742, USA}
\author{Thomas J. Plunkett}
\affiliation{School of Natural Sciences,
University of Tasmania,
Private Bag 37 Hobart, Tasmania, 7001, Australia}

\author{Jean-Baptiste Marquette}
\affiliation{Laboratoire d'astrophysique de Bordeaux, Univ. Bordeaux, CNRS, B18N, alle Geoffroy Saint-Hilaire, 33615 Pessac, France}

\affiliation{Sorbonne Universit\'e, CNRS, Institut d'Astrophysique de Paris, IAP, F-75014, Paris, France}

\correspondingauthor{Natalia E. Rektsini}
\email{efstathia.rektsini@utas.edu.au}

\begin{abstract}

We revisit the planetary microlensing event OGLE-2013-BLG-0132/MOA-2013-BLG-148 using Keck adaptive optics imaging in 2013 with NIRC2 and in 2020, 7.4 years after the event, with OSIRIS. The 2020 observations yield a source and lens separation of $ 56.91 \pm 0.29$ mas, which provides us with a precise measurement of the heliocentric proper motion of the event $\mubold_{rel,hel} = 7.695 \pm 0.039$ mas $yr^{-1}$. We measured the magnitude of the lens in K-band as $K_{lens} = 18.69 \pm 0.04 $. Using these constraints, we refit the microlensing light curve and undertake a full reanalysis of the event parameters including the microlensing parallax $\pi_{E}$ and the distance to the source D$_S$.
We confirm the results obtained in the initial study by \cite{Mroz_2017} and improve significantly upon the accuracy of the physical parameters. The system is an M dwarf of $0.495 \pm 0.054$ $M_\odot$ orbited by a cold, Saturn-mass planet of $0.26 \pm 0.028$ $M_{Jup}$ at  projected separation $r_{\perp}$ = 3.14 $\pm$ 0.28 AU. This work confirms that the planetary system is at a distance of 3.48 $\pm$ 0.36 kpc, which places it in the Galactic disk and not the Galactic bulge.

\end{abstract}

\keywords{Gravitational microlensing --- Adaptive optics --- M-dwarfs}

\section{Introduction}
\label{sec:intro}

Gravitational microlensing is a unique method to discover planets down to the mass of Mars around an unbiased sample of stellar types throughout our Galaxy \citep{gaudi2012}. Its maximum sensitivity is close to the snow line where, according to the core accretion theory, giant planets are mostly formed \citep{1993ARA&A..31..129L,Ida2004ApJ...616..567I}. Furthermore, it can provide information on cold exoplanet demographics throughout the Milky Way and place constraints on planetary formation scenarios \citep[eg.,][]{Suzuki2018AJ....155..263S}. Knowledge of the physical parameters of microlensing events can shed light on the very low planetary mass regions on wide orbits which are technically difficult to access with other detection methods.

The basic microlensing light curve provides precise measurements of the planet-to-host mass ratio and projected separations in units of the Einstein ring
radius. However, additional constraints (e.g.,  finite-source effects, microlensing parallax) and Bayesian analysis are needed to derive the physical parameters, such as the absolute mass and the semi-major axis of the planetary system. If no microlensing parallax constraint can be used, these
physical parameters are often known only to a precision of $\sim 50 \% $ or worse.
 Previous works \citep{Bhattacharya_2018,Vandorou_2020,Bhattacharya_2021,Terry_2021,Batista2015ApJ,2015ApJ...808..169B,Bennett_2020,Blackman2021Natur.598..272B,Ranc2015A&A...580A.125R} have shown that adaptive optics (AO) follow up observations made in the decade following the microlensing event can be used to measure the source-lens flux ratio and separation. This can be translated into a mass-distance relation used to define the physical parameters of the planetary system. Furthermore, when the source and lens are resolved, it is possible to constrain the amplitude and direction of the relative source-lens
proper motion. This gives additional constraints that often help to derive masses and projected separation
to $ \sim 10\%$ precision or better.

In this work we use Keck high angular resolution images, obtained using both NIRC2 and OSIRIS cameras, in order to constrain the planetary mass and distance of OGLE-2013-BLG-0132. Previously \citet{Mroz_2017} has shown that this event can be described by a gas giant planet orbiting an M-dwarf host star beyond the snow line.
M-dwarfs are the most abundant type of star in the galaxy \citep{Winters2015AJ....149....5W},  but the occurrence of gas giant planets orbiting these type of stars is very low.

Core accretion theory \citep{Laughlin_2004, Kennedy_2008} predicts that gas giant planets are expected to be rare around low-mass host stars because they form from a runaway process resulting in the rapid accretion of cold gas onto a planetary core \citep{POLLACK199662, Ida2004ApJ...616..567I}. This means that Jovian and sub-Jovian planet formation requires high solid surface density of the stellar disk and, as a consequence, rapid timescales. This prediction is borne out in planetary population synthesis models treating the host star, disk, and planetesimal accretion self-consistently in N-body simulations \citep[e.g.,][]{Burn2021}, where the gas giant frequency diminishes with decreasing stellar mass and is expected to be very low for M$_* \lesssim 0.5$ M$_{\odot}$. Considering a realistic disk-to-star mass ratio for M-dwarfs would imply that the disks around this type of stars are expected to have difficulty exceeding the threshold density for giant planet formation \citep{Burn2021}. 

Furthermore, the core accretion disk theory predicts a desert area for planets with M$_{p}$ from 10 to 100 M$_{\Earth}$ for orbital distances less than 3 AU.  OGLE-2013-BLG-0132 falls into this intermediate stage with a planet mass of 82.6 M$_{\Earth}$ and projected separation of 3.14 $\pm$ 0.28 AU. The discovery of Jovian and sub-Jovian planets orbiting M-dwarf stars beyond the snow line could imply that the distribution of giant planets is similar for host stars of 0.5 M$_{\odot}$ and for 1.0 M$_{\odot}$ but the number of giant planets is larger for larger host star masses.

Until today there have been 9 confirmed microlensing cases of giant planets orbiting host stars with masses $\lesssim$ 0.6 $M_{\odot}$ that can support this idea. The list of these planets and of the papers confirming their mass measurements is presented in Table \ref{tab:list_planet}. All of the masses of these planets have been determined by our Keck follow-up observations (fourth column).  From the confirmed detections presented here, OGLE-2006-BLG-109Lc and MOA-2009-BLB-319Lb have similar planet and host star masses to OGLE-2013-BLG-0132 while OGLE-2005-BLG-071Lb and OGLE-2003-BLG-235Lb represent Super-Jupiter planets orbiting M-dwarfs. 

\begin{deluxetable*}{cccc}[htb!]
\tablewidth{20cm}
\tablecaption{List of microlensing planets orbiting M-dwarfs, with mass measurements indicating that they are above Neptune's mass.}
\label{tab:list_planet}
\tablehead{
\colhead{Planet name} & \colhead{Planet mass ($M_{Jup}$)} & \colhead{Host star mass ($M_{\odot}$)} & \colhead{Reference}}
\startdata
    OGLE-2005-BLG-071Lb & 3.27 $\pm$ 0.32 & 0.426 $\pm$ 0.037  &   \citep{Udalski_2005,Dong_2009,Bennett_2020}\\
    OGLE-2006-BLG-109Lb & 0.73  $\pm$ 0.06 & 0.51$\pm$ 0.05 & \citep{gaudi2008planet,Bennett_2010_109Lbc}\\
    OGLE-2006-BLG-109Lc & 0.27 $\pm$ 0.02 & 0.51$\pm$ 0.05 &  \citep{gaudi2008planet,Bennett_2010_109Lbc}\\
    OGLE-2007-BLG-349L(AB)c & 0.25 $\pm$ 0.04 & 0.41 $\pm$ 0.07,0.30 $\pm$ 0.07 &  \citep{bennett2016first}\\
    MOA-2008-BLG-379Lb & 3.64 $\pm$ 0.51 & 0.519$\pm$ 0.063 & \citep{Suzuki_2014, bennett_mb08379}\\
    MOA-2009-BLB-319Lb & 0.212 $\pm$ 0.20 & 0.524 $\pm$ 0.048 & \citep{Miyake_2011,Terry_2021}\\
    OGLE-2003-BLG-235Lb & 2.34 $\pm$ 0.43 & 0.56 $\pm$ 0.06 & \citep{Bond_2004,bhattacharya2023confirmation}\\
    MOA-2010-BLG-117Lb & 0.54 $\pm$ 0.10 & 0.58 $\pm$ 0.11 & \citep{bennett2018first}\\
    OGLE-2012-BLG-0950Lb & 0.123 $\pm$ 0.025 & 0.58 $\pm$ 0.04 & \citep{Koshimoto_2017,Bhattacharya_2018}\\
     OGLE-2013-BLG-0132Lb &  0.260 $\pm$ 0.028 & 0.495 $\pm$ 0.054 & \citep{Mroz_2017} \\
      \hline
\enddata
\end{deluxetable*}

Here we combine the microlensing light curve model and the constraints from adaptive optics to acquire a precise mass measurement of the planet-host star system and confirm the event to be in the list of gas giant exoplanets with M-dwarf host stars. Finally, we discuss the significance of the host star mass dependence for the exoplanet formation models.

This paper is organised as follows; First in Section \ref{sec:previous} we discuss the discovery and previous work on OGLE-2013-BLG-0132Lb. In Section \ref{sec:imaging} we describe our Keck AO high angular resolution images and the methods used for image calibration and photometry. We then detail our lens-source relative proper motion and flux ratio measurements. In Section \ref{sec:lc} we perform a Markov Chain Monte Carlo fit to the updated light curve data of the event using the AO constraints and present the best-fit model. In Section \ref{sec:mass-distance} we present the physical parameters of the planetary system. Finally, we discuss our results and conclude the paper in Section \ref{discussion}.

\section{The microlensing event OGLE-2013-BLG-0132} \label{sec:previous}
OGLE-2013-BLG-0132 was discovered and announced by the Optical Gravitational Lensing Experiment (OGLE) Early Warning System (\cite{udalski1994}, \cite{udalski2004optical}) on March 3, 2013. It was also discovered independently as MOA-2013-BLG-148 by the Microlensing Observations in Astrophysics (MOA) collaboration \citep{Bond2001}.  The equatorial coordinates of the event are R.A.$ = 17^h59^m03^s.51 $, dec.\ $= -28^{\circ}25^{'}15{''}.7$ (J2000.0) and the Galactic coordinates  are $l = 1\fdg 944$, $b=-2\fdg275$.

The analysis of the event by \cite{Mroz_2017} yield a Saturn-mass planet orbiting an M-dwarf. Due to the faintness of the source and the short time scale of the event, they obtained only an upper limit on the parallax magnitude $\piEbold \le 1.4$. They derive the angular Einstein radius $\theta_E = 0.81 \pm 0.12$ mas and the Einstein time $t_E = 36.99 \pm 0.77$ days. This  led to a fairly high proper motion of $\mubold_{rel,hel} = 8.0 \pm 1.3$ mas/yr. This value implies a 60 mas source-lens separation after 7.5 years, comparable to the expected FWHM, in the best conditions, with Keck Adaptive Optics systems. This makes OGLE-2013-BLG-0132 a very good candidate for high angular resolution imaging follow up.

In \cite{Mroz_2017} the light curve model gives a planet-star mass ratio of $q =(5.15 \pm 0.28)\times 10^{-4}$. Using a Bayesian analysis assuming that host stars of all masses are equally likely to host a planet of this mass ratio, they estimate the planet and host star masses respectively  to be $ m_p = 0.29^{+0.16}_{-0.13}$ $M_{Jup}$ and $ M = 0.54^{+0.30}_{-0.23}$ $M_\odot$, making the host star an M-dwarf. They performed a grid search on three microlensing parameters $(q,s,\alpha)$, where q is the planet-star mass ratio, s is the projected separation and $\alpha$ is the angle of the source
trajectory with respect to the lens axis. The source angular radius is considered fixed, using a sequential least squares algorithm \citep{kraft}. The source flux magnification was calculated using the ray-shooting method \citep{schneider1986two}, considering the point-source approximation far from the caustic crossings and hexadecapole approximation at intermediate distances \citep{Gould_2008}.

In \cite{Mroz_2017} the estimated brightness of the source at baseline is I$_S$ = 19.37 $\pm$ 0.03 and the color (V$-$I)$_S$ = 1.79 $\pm$ 0.04.  They also measured the red clump centroid on a color-magnitude diagram giving I$_{RC}$ = 15.62 and  (V$-$I)$_{RC}$ = 2.07. Assuming that the source is affected by the same amount of extinction as the red clump stars from the field (\cite{bensby2011chemical}, \cite{nataf2013reddening}), they derived the dereddened color and brightness of the source star: (V$-$I)$_{S,0}$ = 0.78 $\pm$ 0.04 and I$_{S,0}$ = 18.11 $\pm$ 0.20.

Our KECK AO observations are in K-band so we need to transform the $I_S$ magnitude into K magnitude in order to compare our AO results with the fitting model. We choose to use the method of \cite{surot2020mapping} for calculating the (J$-$K$_S$) extinction for the (l,b) galactic coordinates of the event. \cite{surot2020mapping} provide a direct high-resolution (2 arcmin to $\sim$ 10 arcsec) color excess map for the VVV bulge area in (J$-$K$_S$) color, so by using their method we reduce the possible error propagation caused by color-color relations \cite{Bessell_1988}. We find E(J$-$K$_S$) = 0.336 $\pm$ 0.015 for (l,b) = ($1\fdg 944$, $-2\fdg275$). We define the A$_K$ extinction along the line of sight as the total extinction up to the Galactic Center. We use the de-reddened red clump magnitudes of \cite{nishiyama2009interstellar} and obtain A$_J$/A$_K$ = 3.02 which leads to
E(J - K$_S$) = 2.02A$_K$ and finally A$_K$ = 0.181 $\pm$ 0.007.

Finally, we predict the  source magnitude in K-band to be:
\begin{equation}
K_{source} =  V_{S,0} - (V-K)_{S,0} + A_K = 17.35 \pm 0.20.
\end{equation}
As we show in paragraph \ref{sec:flux}, our AO observations in K band confirm this source magnitude. A summary of the color and extinction values are given in Table \ref{tab:flux} and Table \ref{tab:red} respectively.

\begin{deluxetable}{ccc}[htb!]
\tablewidth{20cm}
\tablecaption{  Adopted Extinction and Reddening
Values to the Source}
\label{tab:red}
\tablehead{
\colhead{Parameter} & \colhead{value}}
\startdata
      $A_I$   & 1.26 $\pm$ 0.012 \\
      $A_K$   & 0.181  $\pm$ 0.007\\
      $E(J - K_S)$   & 0.336 $\pm$ 0.015 \\
      \hline
\enddata
\end{deluxetable}

\section{High resolution-imaging follow up} \label{sec:imaging}
\subsection{Analysis of the 2013 NIRC2 Images} \label{sec:2013}
We obtained JHK observations of the target OGLE-2013-BLG-0132 with the NIRC2 instrument and the wide camera (covering a field of 40 arcsec) on Keck-II in July 2013, five months after the peak of the microlensing event. The K-band images had a point spread function (PSF) full width at half maximum (FWHM) of 90 mas.
These observations are used to obtain a calibrated flux measurement at the position of the source, since source and lens are not expected to be resolved that early after the microlensing event.
  We dark subtracted and flat-fielded the images following standard procedures, and we stack the images using SWARP \citep{Bertin}.  We then used the GAIA  catalogue to refine the astrometry of the stacked frames. Finally, using TOPCAT \citep{shopbell2005astronomical}, we cross-identified the catalogues of our re-analysis of the VISTA 4m telescope VVV survey \citep{minniti2010vista} with the KECK sources measured with the SExtractor program \citep{bertin1996sextractor}. The procedures are described in detail in \cite{beaulieu2016revisiting} and \cite{beaulieu2018combining}.

In the 40 arcsec NIRC2 field of view we cross-identified 70 stars also measured in the VVV catalogues. We  then calibrate the KECK frame and estimate that we have an error of the zero point at
$ 2.0\% $ (systematics). We finally provide our measurement of calibrated magnitudes of the source+blended light:
\begin{center}
\begin{align}
\begin{split}
 J_{2013} &= 17.85 \pm 0.05 \\
 H_{2013} &= 17.22 \pm 0.04 \\
 K_{2013} &= 17.05 \pm 0.04 \\
\end{split}
\end{align}
\end{center}

\subsection{Analysis of the 2020 OSIRIS Images} \label{sec:2020}

The second set of observations for the target took place on August 17 and 18, 2020 using the OSIRIS imager on Keck-I. These data were taken using the $K_{p}$ filter and had an average FWHM of 56 mas. The pixel scale of the OSIRIS camera is 9.96 mas/pixel.

We have obtained 25 $K_p$  science images with an individual exposure time of 60 seconds, with 5 dithered positions with an amplitude of 1 arcsec, 80 flat-field frames, 30 dark, and 10 sky frames (60 seconds). We used the Keck AO Imaging (\texttt{KAI})\footnote{https://doi.org/10.5281/zenodo.6677744} data  reduction pipeline \citep{jessica_lu_2022_6522913} to correct dark current, flat-fielding,  instrumental aberrations of the OSIRIS camera and the sky. This tool performs corrections to the 25 science images using the dark, flat-field and sky frames and stacks them into a single master science frame. We then perform the astrometry calibration using the GAIA catalogue as performed for the 2013 NIRC2 images. The combined science frame is presented in Figure \ref{fig:s-l}.
In order to obtain precise positions of the source and lens in the 2020 observations we need to construct an empirical PSF fit for each star individually. We use the methods shown in \cite{Bhattacharya_2018} and \cite{Bhattacharya_2021}, starting with the PSF fitting routine of the DAOPHOT-II package \citep{Stetson_1987}. Once we generate an empirical PSF model we fit it to both source and lens. To do that we fit a two-star PSF to the target using DAOPHOT. The residuals from this method are shown on the right side of Figure \ref{fig:s-l}.

We now have a first guess about the pixel positions of source and lens and their empirical magnitudes, but DAOPHOT does not produce a probability distribution of all possible solutions for our target. It also does not report error bars for the positions of the two stars which means that we can not calculate the precision of the source-lens relative proper motion. For this reason we use a modified version of the original DAOPHOT package \citep{Terry_2021} which contains a supplementary routine that uses the Markov Chain Monte Carlo (MCMC) method to produce a probability distribution for the source-lens parameter space. This parameter space contains six parameters, the $x$, $y$ pixel positions of source and lens, the total flux and the total flux ratio of the two stars. Finally, the routine calculates the $\chi^2$ of each possible solution and returns as a best-fit solution the parameter set with the minimum   $\chi^2$ value.

The quality of AO images are affected by the Strehl ratio. Variability in the atmospheric conditions during observations means that the Strehl ratio and the PSF full width at half maximum values will vary from image to image. Therefore, producing a single master science frame may contain significant imperfections, due to one or more images, that will be included in our PSF model and MCMC results. For this reason we use the jackknife routine \citep{Tukey1958} implemented in the KAI\_Jackknife data reduction pipeline as described in \cite{Bhattacharya_2021}. Using this package we analyze a collection of $N=25$ science images and produce $N$ images of $N-1$ stacked science images. This method helps us detect possible problematic frames and also offers error bars that include the uncertainties of the PSF variations. Finally, we perform the $DAOPHOT\_MCMC $ routine analysis in all 25 jackknife frames. We do that using the same reference stars and magnitude of the target for each frame. We obtain best-fit values and errors from each MCMC and calculate the Jackknife error. Our final uncertainties are the jackknife and MCMC errors added in quadrature as presented in Table \ref{tab:table1}.

\subsection{Resolving source and lens \label{sec:sf}}

\cite{Mroz_2017} gave a source-lens heliocentric relative proper motion of $\mubold_{rel} = 8.0 \pm 1.3$ mas~yr$^{-1}$. We therefore expected a source and lens separation of $\sim 60$ mas in 2020, which is comparable to the average PSF FWHM of the OSIRIS images.

\begin{figure*}
\plotone{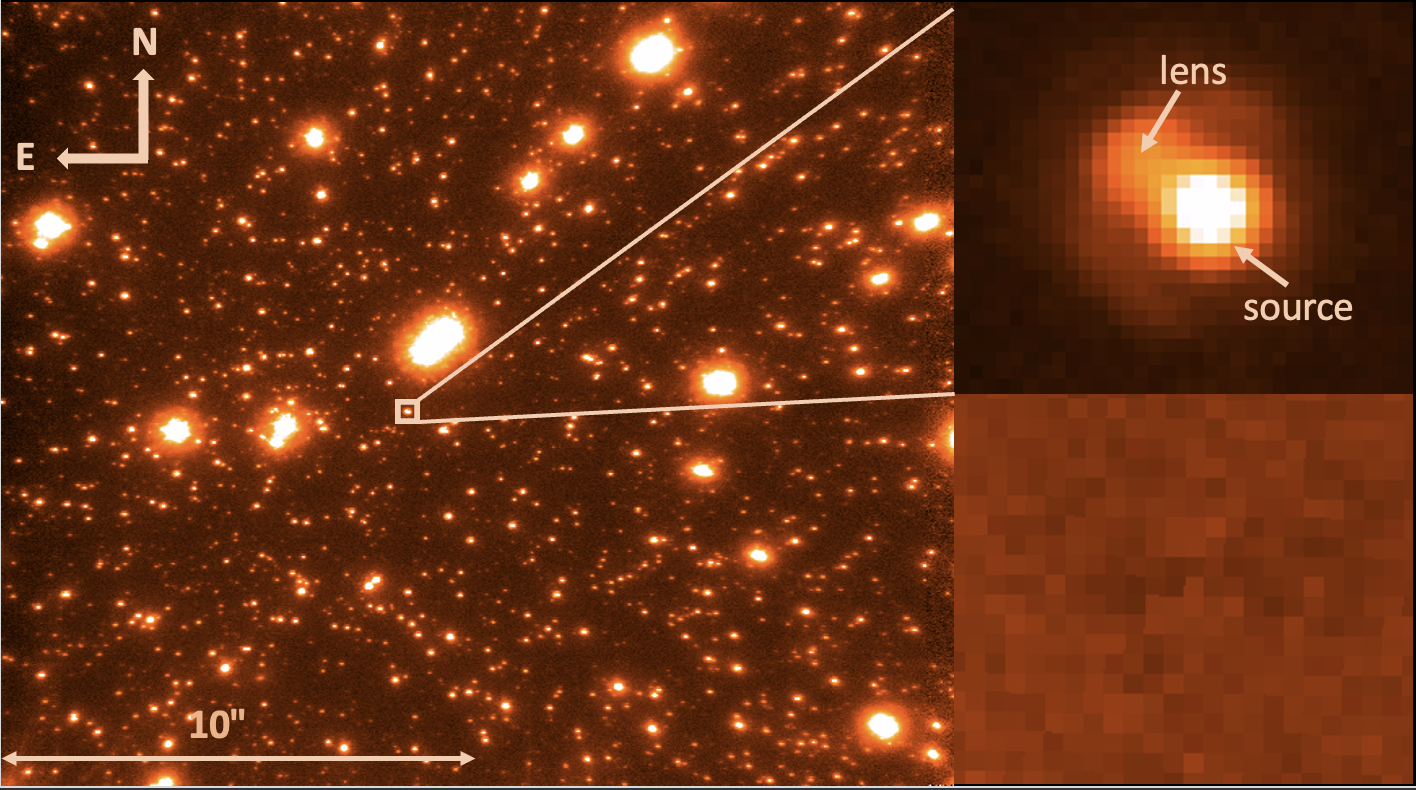}
\caption{Left panel: jack-knife stack of 24 (60 seconds) frames of the 2020 Keck OSIRIS $K_p$ band follow up observation. Upper right panel: close-up { \bf (2.5"$\times$ 2.5")} frame of the source and lens. Lower right panel: close-up { \bf (2.5"$\times$ 2.5")} of the two-star PSF fit residual using DAOPHOT.
\label{fig:s-l}}
\end{figure*}

\begin{figure}
\plotone{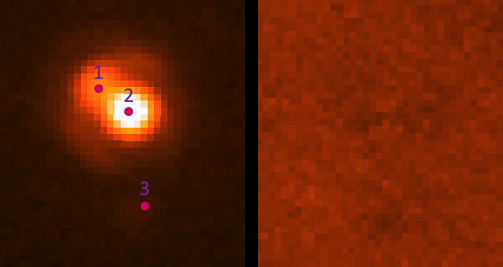}
\caption{Left panel: Close-up frame (400x420 mas) of the three-star group selection by DAOPHOT. Right panel: residual of the three-star PSF fit.
\label{fig:3stars}}
\end{figure}

A visual inspection of the KAI\_Jackknife combined images showed two stars at the position of the microlensing event (Figure \ref{fig:s-l}). Thus, source and lens are resolved enough for the DAOPHOT routine to be able to identify them as two separate stars. Due to the crowded image the routine identifies a nearby third  star in a distance of 16 pixels from the source-lens center. We compare the parameter space results for a two-star and three-star PSF fit using the basic DAOPHOT routine.
In Table \ref{tab:daophtot} we present the results of our DAOPHOT analysis for the two-star and three-star PSF model including the pixel coordinates and instrumental magnitudes for each star component. We measure the separation and the total flux ratio F=F1/(F1+F2) between star1 and star2 for each model case.  Our three-component frame and the three-star residual is shown in Figure \ref{fig:3stars}.
We find the third component to have a separation of $\sim$ 154 mas from the first star and $\sim$196 mas from the second star. Furthermore it has a magnitude difference of more than 5 magnitudes from the bright star (star1) and almost 3 magnitudes from the faint star (star2). Its large separation from the other components and its faintness makes it an unsuitable candidate for either the source or the lens in this study. In addition, the difference in the source-lens separation between the 2-star and 3-star models is 0.565 mas thus, the difference between the two models is less than the relative proper motion error bar  derived from \cite{Mroz_2017} . Since our results for the source-lens separation and flux ratio are not significantly affected by the inclusion of the third star in the group we decided to maintain the 2-star model for the rest of this work.

\begin{deluxetable*}{lccccccc}[htb!]
\tablewidth{20cm}
\tablecaption{DAOPHOT results for the 2-star and 3-star PSF fit for the 2020 OSIRIS images \label{tab:daophtot}}
\tablehead{
\colhead{Model} & \colhead{Component}& \colhead{Coordinates} & \colhead{mag} & \colhead{s (mas)} & \colhead{F} & \colhead{$\chi^2$/ dof} }
\startdata
    {       }& $Star_1$ &    [1210.92,1420.87] &    12.867 \\
    {       }& $Star_2$ &    [1206.48,1424.46] &    14.261 \\
    2-star model& {       } &    {       } &    {       } & 56.885  &     0.7806  &     544.12/621 \\
    \hline
    {       }& $Star_1$ &    [1210.93,1420.88] &    12.729 \\
    {       }& $Star_2$ &    [1206.48,1424.47] &    14.146 \\
    {       }& $Star_3$ &    [1211.48,1405.44] &    17.269 \\
    {3-star model}& {       } &    {       } &    {       } &56.798  &     0.7781 &     259.42/618 \\
    \hline
\enddata
\tablecomments{The pixel coordinates, instrumental magnitude, separation, total flux ratio and $\chi^2$ values for the two-star and three-star PSF fitting models. The values show the results of the basic DAOPHOT routine for only one image combination of N-1 image frames. The separation and total flux ratio differences between the two models are inside the value's uncertainties derived by \cite{Mroz_2017}.}
\end{deluxetable*}

\subsection{Flux ratio} \label{sec:flux}
We start by measuring the magnitude in $K_p$ band for the source and lens separately. As shown in Section \ref{sec:2013} the source-lens blending magnitude measured by the KECK 2013 images is $ K_{Keck} = 17.05 \pm 0.04$. The $DAOPHOT\_MCMC$ analysis of the 2020 OSIRIS images yield a total flux ratio for the two stars to be F = $0.778 \pm 0.001$. By combining the previous results with the absolute magnitude and flux ratio equations we calculate the source and lens absolute magnitudes:
\begin{equation}
    \frac{f_L}{f_S}= \frac{1-F}{F} ,
\end{equation}
\begin{equation}
    K_{Keck,L} - K_{Keck,S} = -2.5 \log_{10}(\frac{f_L}{f_S}) 
\end{equation}
and
\begin{equation}
    K_{Keck} = -2.5 \log_{10}(10^{-0.4 K_{Keck,S}} + 10^{-0.4 K_{Keck,L}})
\end{equation}
 The results we obtain from these Keck observations and from the light curve best fit model imply that the brightest star corresponds to the source. In addition, in Table \ref{tab:flux} we compare our brightest star magnitude with the source star magnitude deduced by \cite{Mroz_2017} and show that the two magnitudes are in good agreement. This is a strong proof of the source star identification    
 and as a result we have: \\
\\
$ f_L/f_S = 0.2848 \pm 0.0019$
\\
$ K_{Keck,L} = 18.69 \pm 0.04$
\\
$ K_{Keck,S} = 17.32 \pm 0.04$ \\

\subsection{Lens-source relative proper motion} \label{sec:murel}

There exist two methods for measuring the source-lens relative proper motion. One method would be to use the finite source effects on the light curve, if these are detected, and a color-magnitude diagram \citep{Boyajian_2014}. Dividing the angular source radius $\theta_*$ by the source radius crossing time $t_*$ leads to the measurement of the geocentric relative proper motion $\mubold_{rel,geo} = \theta_*/t_*$. Using the values from the light curve fitting model presented in the Section \ref{sec:lc} we find that $\mubold_{rel,geo} = 7.684 \pm 0.881$ mas~yr$^{-1}$. This measurement of the relative proper motion yields an angular Einstein radius of $\theta_E = 0.78 \pm 0.09$ mas.

Another way to deduce the relative proper motion and then the angular Einstein radius, with higher precision, is the use of AO follow-up images. Here the 2020 Osiris images give us a separation s = $56.911 \pm 0.290$ mas, 7.403 years after the peak of the microlensing event. We convert it into an heliocentric relative proper motion $\mubold_{rel,hel}$ = 7.688 $\pm$ 0.039 mas~yr$^{-1}$. In Table \ref{tab:table1} we present our final results from the 2020 Osiris images. 

Since the light curve model uses a geocentric reference frame, relative proper motion must be expressed  in the inertial geocentric frame. We use the relation given by \cite{Dong_2009} in order to convert  the heliocentric relative proper motion into geocentric relative proper motion $\mubold_{rel,geo}$ :
\begin{equation}
    \mubold_{rel,geo}  = \mubold_{rel,helio} - \Delta\mubold
\end{equation}
where
\begin{equation}
    \Delta\mubold = \frac{\pi_{rel} V_{\oplus,\perp}}{AU} = (\frac{1}{D_L} - \frac{1}{D_S})\Vbold
\end{equation}

The $\Vbold$ represents the velocity of the Earth projected on the sky at the R.A., dec coordinates at the peak of the microlensing event  $  (17^{h}59^{m}03^{s}.689 $, $-28^{\circ}25^{'}16{''}.29)$. The velocity is expressed in north and east coordinates :
\begin{equation}
    \Vbold = (2.61, -1.45)\,\, \mathrm{km/sec}
\end{equation}

Calculation of the relative distance of the source-lens $\pi_{rel}$ 
demands the definition of the distance of the lens at the time of the event. 
In Section \ref{fig:lc_fit} we use the high angular resolution data to constrain the light curve fitting models that provide microlensing parallax values that are in agreement with the AO results for the source-lens relative proper motion and the flux of the lens. This method leads to $\pi_{rel}$ = 0.423 $\pm$0.030 (kpc)$^{-1}$ which yields a geocentric relative proper motion 7.594 $\pm$ 0.052 mas/yr. Finally, using this method we deduce an angular Einstein radius of $\theta_E = 0.785 \pm 0.017$ mas.

\begin{deluxetable*}{ccccc}[htb!]
\tablewidth{20cm}
\tablecaption{$DAOPHOT\_MCMC $ and Jackknife Best-fit results for the 2020 Osiris images}
\label{tab:table1}
\tablehead{
\colhead{Parameter} & \colhead{Median} & \colhead{MCMC rms} & \colhead{Jackknife rms} & \colhead{MCMC + JK rms}}
\startdata
    Separation (mas) & 56.911  & $\pm$ 0.232  &    $\pm$      0.174
    &    $\pm$ 0.290\\
      $\mu_{rel,HE}$(mas~yr$^{-1}$) & -4.824 & $\pm$ 0.026 &  $\pm$ 0.029
      & $\pm$ 0.039\\
      $\mu_{rel,HN}$(mas~yr$^{-1}$) & 5.985 & $\pm$ 0.029 &  $\pm$ 0.032
      & $\pm$ 0.043\\
      $\mubold_{rel,helio}$(mas~yr$^{-1}$) & 7.695 & $\pm$ 0.031 &  $\pm$ 0.024
      & $\pm$ 0.039\\
      $\mubold_{rel,geo}$(mas~yr$^{-1}$) & 7.594 & $\pm$ 0.034 &  $\pm$ 0.040
      & $\pm$ 0.052\\
      flux ratio   & 0.2848 & $\pm$ 0.0028 & $\pm$ 0.0019
      &   $\pm$ 0.003 \\
      \hline
\enddata
\end{deluxetable*}

\section{Light curve fitting} \label{sec:lc}

In this work we use a modified version of the imaged-centered ray shooting light curve modeling code
of \cite{bennett96} and \cite{bennett-himag}, which now
goes by the name, \texttt{eesunhong}\footnote{\url{https://github.com/golmschenk/eesunhong}}, 
in honor of the original co-author of the code \citep{rhie_phystoday,rhie_obit}. This new version of the
\texttt{eesunhong} code incorporates constraints from Keck AO on lens flux measurements and lens-source relative 
proper motion on the light curve models. This code also includes the microlensing parallax parameters, even when they
are not determined by the light curve, because they are tightly constrained by the relative proper motion measurements
and the lens magnitude. Inclusion of the microlensing parallax can be important because the microlensing parallax 
parameters can influence Einstein radius crossing time and the inferred source star magnitude, as was shown
by \cite{bennett_mb08379}.

There is a complication that comes from using the measured relative proper motion from the Keck AO data to constrain the microlensing 
parallax, because the Keck AO data determines the relative proper motion in the Heliocentric reference frame, $\mubold_{\rm rel,helio}$, while the microlensing parallax vector is parallel to the relative proper motion, $\mubold_{\rm rel,geo}$,
in the inertial geocentric frame that moves with the Earth at the time of the event. This requires that we add the 
source distance, $D_s$ as a model parameter, which we constrain with a prior from a Galactic model \citet{koshimoto_gal_mod}.

This modeling method is explained in more detail in \cite{bennett_mb08379}, and has also been used in the analysis of
OGLE-2016-BLG-1195 \citep{vandorou_ob161195}.

\subsection{Survey Data}
The event OGLE-2013-BLG-0132 was only observed by the OGLE and MOA ground-based photometric surveys \citep{Mroz_2017}. The MOA photometric data contained systematic errors due to the faintness of the source star, weather conditions and also the larger pixel scale than OGLE. For this reason \cite{Mroz_2017} used only a subset  of the MOA dataset within $\pm$10 days of the peak, including the caustics and ignoring the wings. Here we revisit the modeling of the light curve using a re-reduction of MOA data which performs a de-trending process to correct for systematic errors and removes correlations in the data \citep{bond2017lowest}. We were therefore able to use three years of data around the peak of the event.
Finally, \cite{Mroz_2017} mentioned a long-term trend in the OGLE data, which was treated before their light curve fitting. This trend is probably caused by a very bright nearby star that is moving with respect to the target. Since no microlensing parallax had been observed the OGLE data treatment are sufficient for modelling this event.

\subsection{Light Curve Modelling}

The model presented by \citep{Mroz_2017} shows no ambiguity in the light curve parameters and we have no reason to expect significant differences using the re-reduced data. Eesunhong uses high angular resolution results for the source-lens relative proper motion and the flux of the lens as additional constraints in the Markov Chain Monte Carlo analysis. This method ensures that the light curve parameters are consistent with the AO follow-up observations and allows us to fit the microlensing parallax, even when this hasn't been observed/constrained during the event.
We modelled the light curve  of the event  using the image-centred ray-shooting method \citep{Bennett_1996}. We begin by using the original light curve fitting code exploring the parameter space for a binary lens and a single source star (2L1S), using the best-fit results presented in \citep{Mroz_2017} as initial conditions.

There are seven basic parameters that describe the shape of a light curve of a microlensing event. Three of these parameters describe both a single and binary lens model: $t_E$, the Einstein radius crossing time that defines the event's time scale, $t_0$, the time of the minimum approach of the lens center of mass, and $u_0$, the impact parameter relative to the lens center of mass.
When the source star transits a caustic or a cusp, we can measure the fourth parameter $t_{*}$, the source radius crossing time. We use finite source effects for the measurement of the source-lens relative proper motion.

The final three parameters describe the physical parameters of a binary lens system. These are the planet-star mass ratio, $q$, their projected separation, $s$, in Einstein radius units, and the angle between the planet-star separation vector and the source trajectory, $\alpha$. We fit the light curve model by using this set of parameters in order to predict the flux of the event. To do this we fit two additional observational parameters per observing site, the source star flux $F_s$, and the unmagnified blend flux  $F_b$, which might include the lens flux, as well as close neighbour stars. The light curve model is defined as $F(t) = A(t)F_s + F_b$, where F(t) is the flux of the event at time t.

\begin{figure*}[t!]
\begin{center}
\plotone{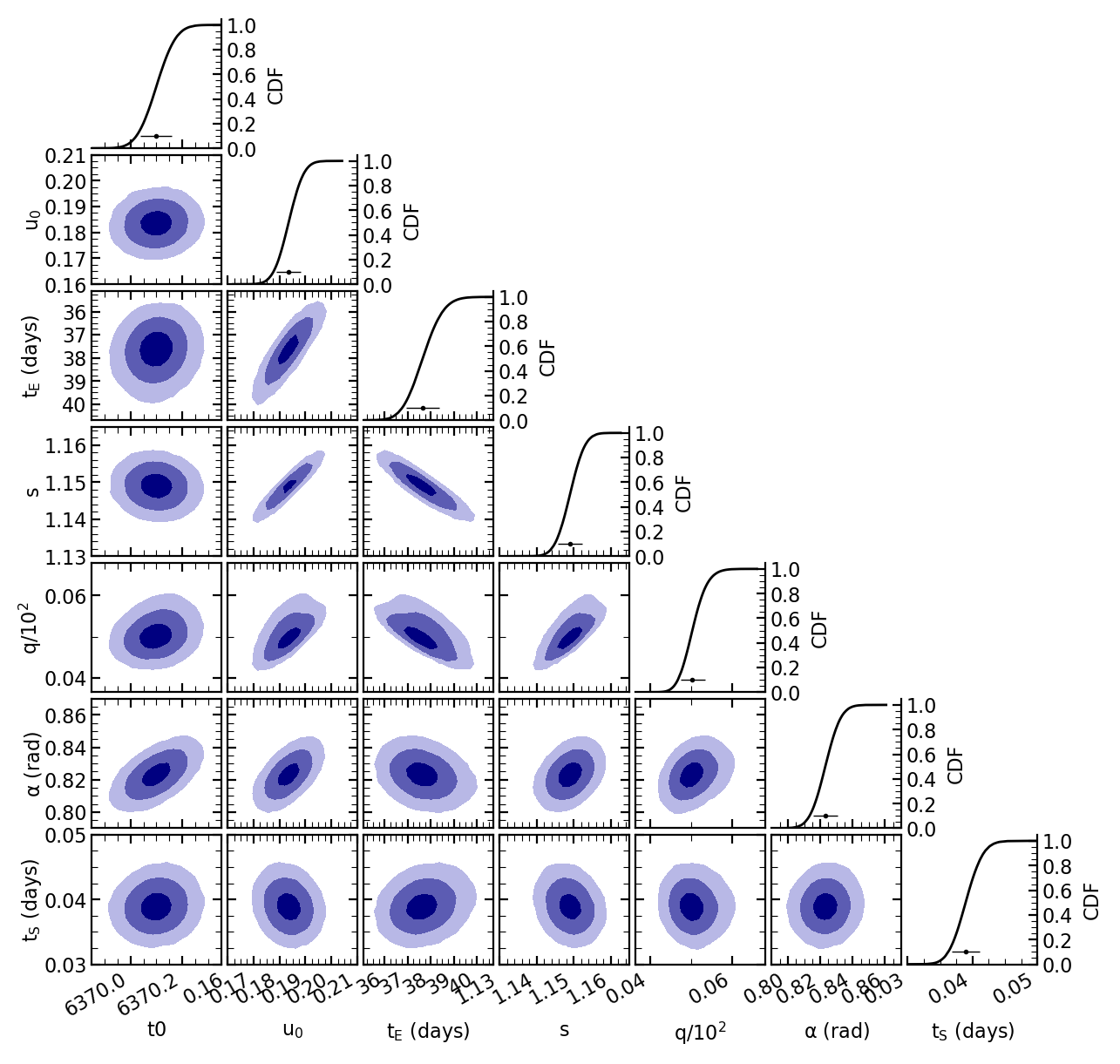}
\caption{The marginalized posterior distributions of the light curve best fit model. In the diagonal we show the one-dimensional cumulative density function of each parameter. The 68.3$\%$ (1$\sigma$), 95.5$\%$ (2$\sigma$) and 99.7$\%$ (3$\sigma$) confidence intervals are represented by dark, median and light violet respectively.
\label{fig:posteriors}}
\end{center}
\end{figure*}

\begin{figure}
\plotone{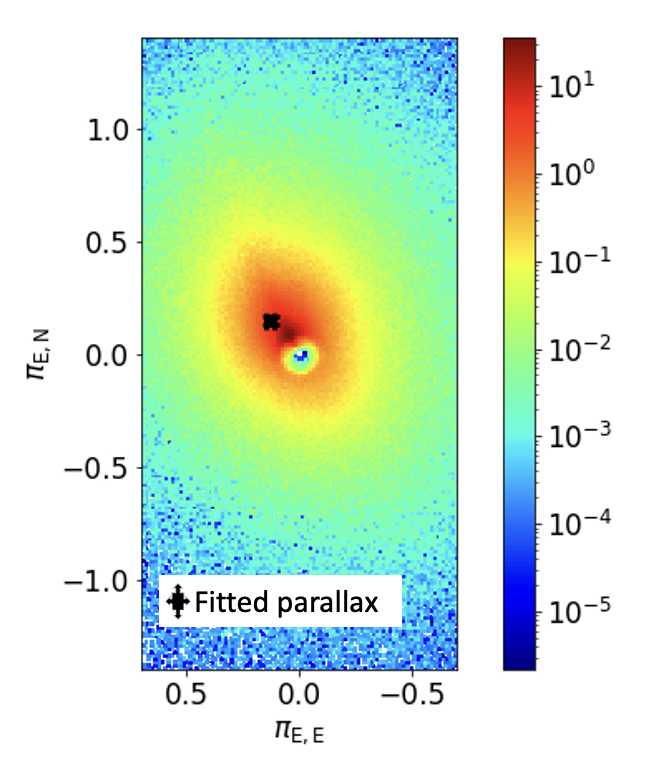}
\caption{Two-dimensional parallax distribution based on the Galactic model (\texttt{genulens}). The color-scale shows the relative probability, the black cross indicates the microlensing parallax predicted using our (AO) Keck constraints.
\label{fig:pie}}
\end{figure}

\begin{figure*}
\plotone{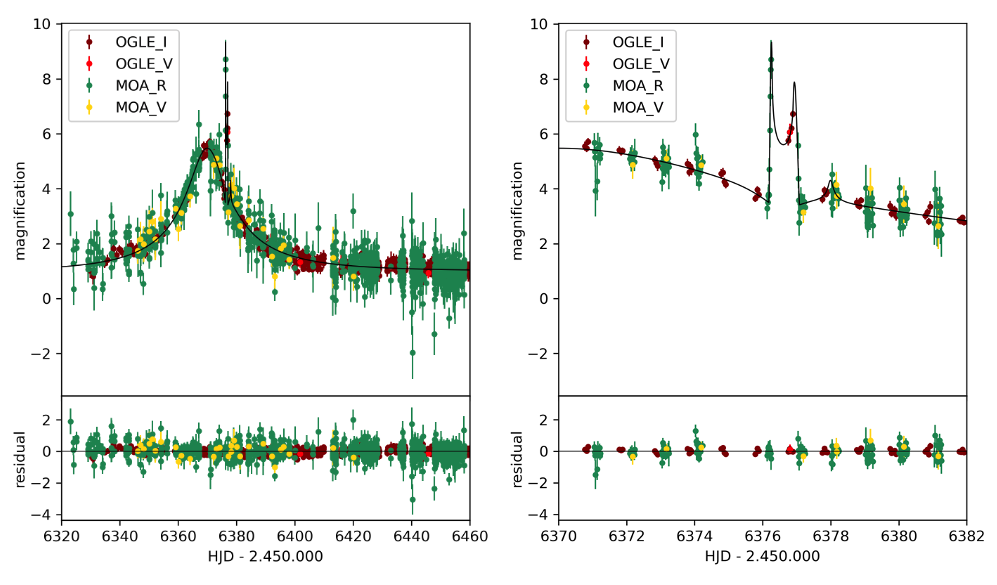}
\caption{Light curve of OGLE-2013-BLG-0132. The right figure shows the enlargement of the caustic-crossing part of the light curve. The best-fit model is indicated by the black curve. The bottom panel shows the residual from the best-fit model and the OGLE and MOA data. The figures were made using the software described in \cite{moana} .
\label{fig:lc_fit}}
\end{figure*}

Once we fit the light curve  we used an MCMC algorithm with a Metropolis Hastings sampler to inspect the posterior distributions of the lens physical parameters as shown in Figure \ref{fig:posteriors}. We then use the mean values of the distributions as initial conditions for our second light curve fitting that contains the source-lens relative proper motion and the lens magnitude found as a prior in Section \ref{sec:sf} from the Keck 2020 follow up images. These two parameters, when defined with high accuracy, place a strong constraint on the microlensing parallax. Figure \ref{fig:lc_fit} shows the light curve best fit and the residual from the MOA and OGLE data during the magnification event. We show that the model describes the planetary anomaly with high precision. In Figure \ref{fig:pie} we show the two-dimensional parallax values that the best fit model yields plotted on the parallax distribution based on the Bayesian analysis used (\texttt{genulens})\footnote{https://doi.org/10.5281/zenodo.6869520}  described in \cite{naoki_koshimoto_2022_6869520,Koshimoto_2021}. The parallax deduced by the Keck constraints matches the high relative probability region predicted by the Galactic model.
As mentioned in Section \ref{sec:murel} the relative proper motion must be in geocentric coordinates for defining $\theta_{\epsilon}$. This means that we need to include the distance to the source as a fitting parameter in our light curve model. For the initial estimate we choose to use the $D_s$ value calculated by the \cite{Koshimoto_2021} Galactic model.
Fitting a parallax distribution that is in agreement with the high angular resolution follow up leads to an additional constraint for the mass and distance of the planetary system.

Our fitting parameters are consistent  with the \cite{Mroz_2017} results with some small differences in the Einstein radius crossing time, mass ratio and the modelled source flux. We have achieved a significant increase in the accuracy in most of our parameters upon the previously published results, especially for $u_0$, s, $t_*$ and $I_s$. We present all the parameters in Table \ref{tab:lc_fit}.

We re-measured the calibrated source color to $(V-I)_S = 1.79 \pm 0.04$ and deduce the de-reddened color and I-band magnitude of the source star as $(V-I)_{S,0} = 0.78 \pm 0.04$ and $ I_{S,0} = 18.01 \pm 0.04 $. Finally, we use the de-reddened values of the source star in combination with the surface brightness relations from \cite{Boyajian_2014} in order to determine the angular source size $\theta_*$ :
\begin{equation}
    log(2\theta_*) = 0.5014 + 0.419(V-I)_{S,0}
    - 0.2 I_{S,0}
\end{equation}
which gives us an angular source size of $\theta_* = 0.80 \pm 0.08$ $\mu as$.

This result is in good agreement with \cite{Mroz_2017} but with improved error bars.

\begin{deluxetable}{ccc}[htb!]
\tablewidth{20cm}
\tablecaption{Source Flux Values}
\label{tab:flux}
\tablehead{
\colhead{Parameter} & \colhead{Mroz+17} & \colhead{this work}}
\startdata
      $I_{S,0}$ & 18.11 $\pm$ 0.20 & 18.01 $\pm$ 0.04\\
      $K_{S,0}$ & 17.365 $\pm$ 0.20 & 17.32 $\pm$ 0.04\\
      \hline
\enddata
\end{deluxetable}

\begin{deluxetable}{cccc}[htb!]
\tablewidth{20cm}
\tablecaption{Light curve best-fit model parameters. We show the MCMC mean values and 1$\sigma$ results for the best-fit obtained using only the light curve data (Column 1), the light curve data and the constraints derived by our 2020 Keck follow-up images (Column 2) and the results presented by \cite{Mroz_2017} in the discovery paper (Column 3).}
\label{tab:lc_fit}
\tablehead{
\colhead{Parameter} & \colhead{MCMC (lc)} &
\colhead{MCMC (lc + AO)} &\colhead{Mroz+17}}
\startdata
      $t_E$ (days) & 37.60 $\pm$ 0.72  &    37.40 $\pm$ 0.26  &36.99$\pm$      0.77\\
      $t_0$(HJD') & 6370.097$\pm$ 0.059 &
      6370.062$\pm$ 0.059 & 6370 $\pm$ 0.064\\
      $u_0$ & 0.183 $\pm$ 0.004 &
      0.185 $\pm$ 0.001 & 0.184 $\pm$ 0.005 \\
      s   & 1.149 $\pm$ 0.003 &
      1.1502 $\pm$ 0.0009 & 1.150 $\pm$ 0.004 \\
      $\alpha$ (rad) & 0.822 $\pm$0.005 &
      0.821 $\pm$0.006 & 0.821 $\pm$ 0.008 \\
      q $\times 10^{-4}$ & 5.00 $\pm$0.27 &
      5.00 $\pm$0.28 & 5.15 $\pm$ 0.28 \\
      $t_*$ (days) & 0.038$\pm$0.002 &
      0.0387$\pm$0.0016 & 0.037 $\pm$ 0.004 \\
      $I_s$ & 19.349$\pm$0.031 &
      19.277$\pm$0.004 & 19.37$\pm$  0.03 \\
      $\pi_{E,E}$ & -- & 0.129 $\pm$ 0.019 &  -- \\
      $\pi_{E,N}$ & -- & 0.145 $\pm$ 0.023 &  -- \\
      $\piEbold$ & -- & 0.195 $\pm$ 0.023 &  $<1.4$ \\
      $D_s$(kpc) & -- & 7.405 $\pm$ 0.710 & --\\
      $\chi^2$ & 9389/9404 & 9384/9401 & 1104/1019 \\
      \hline
\enddata
\end{deluxetable}

\section{Planetary system parameters } \label{sec:mass-distance}

The lens magnitude and lens-source separation we have deduced from our (AO) images permit us to use all three empirical mass-luminosity relations and finally derive a measurement of the mass and distance of the lens.

First, we use a mass-distance relation from resolving the source and lens system  ($\mubold_{rel,geo}$) which constrains the angular Einstein ring radius ($\theta_E$) as shown in \ref{sec:murel}:

\begin{eqnarray}
    M_L = \frac{\theta_E^2}{\kappa \pi_{rel}}
\end{eqnarray}

with $ \pi_{rel} = au ({D_L}^{-1} - {D_S}^{-1}) $ the relative source and lens distance and $\kappa = \frac{4G}{c^2 AU}$ = 8.144 mas$M_{\odot}^{-1}$. $M_L$ is the lens mass, $D_L$ the distance to the lens and $D_S$ the distance to the source  derived by the light curve best fit in Section \ref{sec:lc}.

Fort the second mass-distance relation we use the microlensing parallax expressed by:
\begin{equation}
     \piEbold = \sqrt{\frac{\pi_{rel}}{\kappa M_L}}
\end{equation}

\begin{figure*}
\plotone{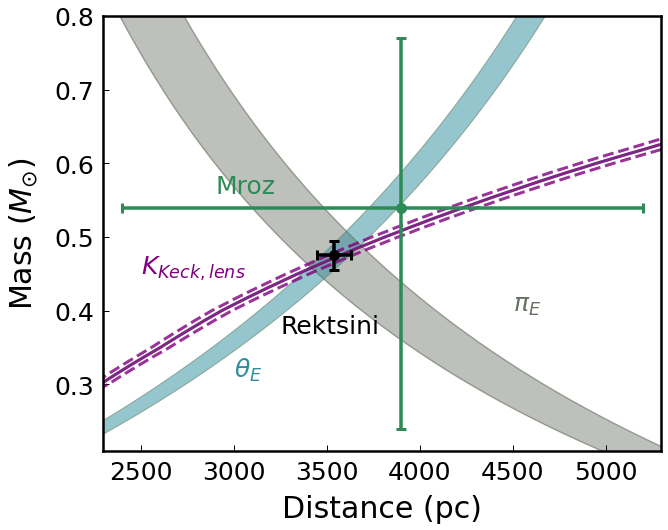}
\caption{Mass-distance estimate for the lens. The purple curve represents the constraint from the K-band lens flux measurement, the seagreen curve shows the Einstein angular radius measurement and the grey curve represents the microlensing parallax calculated using the (AO) constraints. The intersection between the three curves defines the estimated solution of the lens physical parameters.
\label{fig:M-D}}
\end{figure*}

Finally, we proceed by correlating the lens magnitude measured by Keck with a calibrated population of main sequence stars. For this  we use isochrones \citep{girardi2002theoretical} that provide a mass-luminosity function for different ages and metallicities of main sequence stars. We decide to use isochrones for ages 500 Myrs to 6.4 Gyrs and metalicities within the range 0.0 $\le$ [Fe/H] $\le$ +0.2. We finally combine the lens magnitude and the isochrones in order to deduce an independent mass-distance relation : \\
\begin{eqnarray}
    m_L(\lambda) = 10 +5log_{10}(D_L/1 kpc) + A_{K_L}(\lambda) +\\
    \nonumber 
    \eqnum{12}
    M_{isochrone}(\lambda, M_L, age, [Fe/H])
\end{eqnarray}
    
where $m_L$ is the magnitude of the lens, $A_{K_L}$ the extinction to the lens, here in K-band, and $M_{isochrone}$ is the absolute magnitude of the lens star at wavelength $\lambda$. We can determine the distance and the mass of the lens through the intersection of these three relations as shown in figure \ref{fig:M-D}.

$A_{K_L}$ is estimated by considering the source distance and reddening determined above and the distribution of Galactic dust relative to the source and lens distances.
We calculate it as a function of the lens distance $D_L$, given its Galactic coordinates (l,b)= (1\fdg9444, -2\fdg2745). Assuming that the dust in the Milky Way is distributed in an exponential disk in both radius and height \citep{drimmel2001three}, the extinction along any disk sightline can be approximated as:
\begin{equation}
    A_{K_L} = \frac{1- e^{-\vert D_L(sinb) / h_{dust}\vert}}{1- e^{-\vert D_{S}(sinb) / h_{dust}\vert}} A_K
\end{equation}

where $h_{dust}$ is the dust scale height fixed at $ h_{dust} = 0.10 \pm 0.02 $ kpc  and $D_S$ is the distance to the source derived in Section \ref{sec:lc}. We use the $A_K$ extinction value calculated as shown in Section \ref{sec:previous}. This gives us a value for the K-band lens extinction of $A_{K_L} = 0.179$.

\begin{figure*}
\plotone{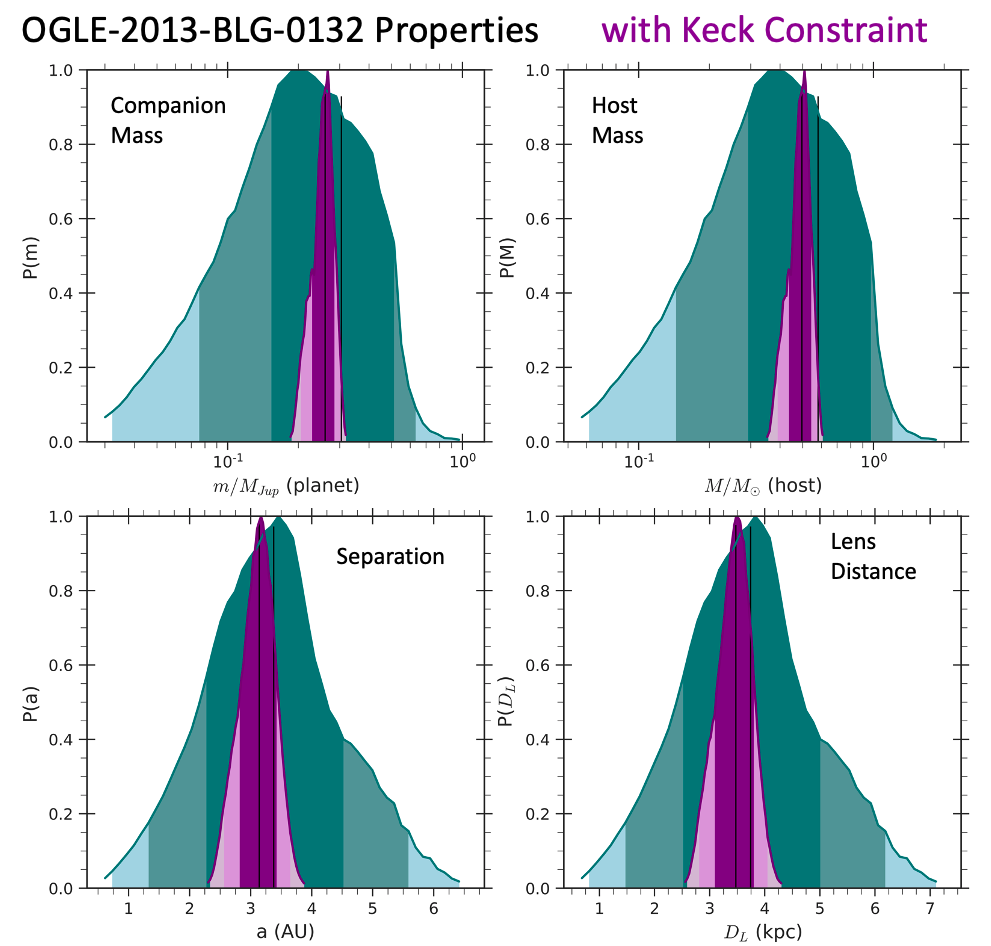}
\caption{Bayesian posterior probability distributions for the planetary companion mass, host mass, their separation, and the distance to the lens system are shown
with only light-curve constraints in blue and with the additional constraints from our Keck follow-up observations in red. The central 68.3$\%$ (1$\sigma$) of the distributions are
shaded in darker colors (dark magenta and dark cyan), and the remaining central 95.4 $\%$ (3$\sigma$) of the distributions are shaded in lighter colors. The vertical black line marks the
median of the probability distribution for the respective parameters. We show that the medians of the Bayesian probability are within 2$\sigma$ of the constrained parameter distributions.
\label{fig:hist}}
\end{figure*}

In Figure \ref{fig:M-D} the isochrone constraint is in purple, the dashed lines indicate the error on the measured lens magnitude, the Einstein angular radius is shown in seagreen and the microlensing parallax constrain in gray.
The result of the combined mass and distance relations is in perfect agreement with the MCMC mean and rms results yield by the light curve model fit with the KECK (AO) constraints (Table \ref{tab:lenspar}) and shown in Figure \ref{fig:hist} in magenta colors. We confirm that the host is an M-dwarf and the planet is a Saturn-mass planet with a projected separation:
\begin{equation}
    r_{\perp} = sD_L\theta_E,
\end{equation}
and we find $r_{\perp} = 3.140 \pm 0.281$ AU.

\begin{deluxetable}{ccc}[htb!]
\tablewidth{20cm}
\tablecaption{Lens Parameters Table}
\label{tab:lenspar}
\tablehead{
\colhead{Parameters} & \colhead{Units} &\colhead{Values and 1$\sigma$}}
\startdata
      $M_h$ & $M_{\odot}$   & 0.495 $\pm$ 0.054  \\
      $M_p$ & $M_{Jup}$   & 0.260 $\pm$ 0.028  \\
      $D_L$ & kpc   & 3.476 $\pm$ 0.357 \\
      $r_{\perp}$ & au & 3.140$\pm$ 0.281\\
      $\theta_E$ & mas & 0.785 $\pm$ 0.017\\
      \hline
\enddata
\end{deluxetable}

\section{Discussion}\label{discussion}
We observed OGLE-2013-BLG-0132 using AO techniques with Keck three months and 7.4 years after the microlensing event. Our 2020 high angular resolution images showed a clear separation between source and lens. This is the sixth microlensing event where the relative positions and flux ratio between source and lens were accurately measured.
We used an MCMC routine of the DAOPHOT package \citep{Terry_2021} and a jackknife routine of KAI  as in \cite{Bhattacharya_2021} and deduced the flux of the lens and a ten times more accurate value for the Einstein angular radius. Our analysis has showed a separation of 56.91 $\pm$ 0.29 mas which yields a $\mubold_{rel,helio}$ = 7.695 $\pm$ 0.039 mas/yr. We confirm the results presented by \cite{Mroz_2017} for the microlensing event OGLE-2013-BLG-0132, then refine them.

As a consequence of the high precision of our AO results we were able to use a modified version of the \cite{bennett96} and \cite{Bennett_2010} process. We fit the event's light curve while constraining the best-fit model using the Einstein angular radius, relative proper motion and source flux in K-band that we deduce from the high angular resolution image analysis. This is the most rigorous way to find light curve parameters able to define the source and lens system without inconsistencies.
Using the AO follow-up constraints we successfully  fit the microlensing parallax and the distance to the source. Our light curve best fit model is in agreement with the previous results. Our fitted microlensing parallax is in agreement with the predicted values from the galactic model \citep{Koshimoto_2021}. Our measurement of the finite source effects and the mass and distance of the planetary system validate the estimates predicted in \cite{Mroz_2017}.

We find the source angular radius to be $\theta_* = 0.80 \pm 0.08$ $\mu as$, at a distance of 7.405 kpc, which means that the source star must have a radius of $\sim$1.27 $R_{\odot}$. We also measured the source brightness in K band to be  $K_{S,0}$ = 17.32 $\pm$ 0.04. This makes the source star a possible early G or late F-type star placed in the Galactic bulge. Finally, our measurements confirm that the OGLE-2013-BLG-0132 event consists of an M-dwarf host star with mass $M_h$ = 0.495 $\pm$ 0.054 $M_{\odot}$ and a Saturn-mass planet with $M_p$ = 0.26 $\pm$ 0.028 $M_{Jup}$ orbiting beyond the snow-line location (2.7 AU) at 3.14 $\pm$ 0.28 AU. With a mass ratio of q = $5\times 10^{-4}$ this system is placed just outside of the planet desert ($1\times 10^{-4}<$q$<4\times 10^{-4}$) predicted by the core accretion theory \citep{Ida2004ApJ...616..567I} and by the existing population synthesis models \citep{Laughlin_2004,mordasini2009extrasolar}.

However, most of these theoretical work consider the planet and host star masses instead of mass ratios. As mention in Section \ref{sec:intro},  \cite{Ida2004ApJ...616..567I} predict the planetary desert for masses between 10 and 100 M$_{\Earth}$ explaining that planets’ masses grow rapidly from 10 to 100 M$_{\Earth}$, the gas giant planets rarely form with asymptotic masses in this intermediate range. \cite[Figure 7][]{Suzuki_2016} indicates that the detection efficiency for planets like OGLE-2005-BLG-071Lb \citep{Bennett_2020} and MOA-2008-BLG-379Lb \citep{bennett_mb08379} with ${\simeq}$ 6.$10^{-3}$ or 7.$10^{-3}$ is about 5 times larger than the detection efficiency for planets with q $\simeq$ 5.$10^{-4}$, like OGLE-2013-BLG-0132 and OGLE-2006-BLG-109Lc \citep{gaudi2008planet,Bennett_2010_109Lbc}. Our results in addition to these other similar results presented in Section \ref{sec:intro} tend to agree with the \cite{Suzuki_2016} conclusion that sub-Saturn mass planets are likely to be 5 times more common than super-Jupiters for early M-dwarfs instead of just the average over all planets detectable by microlensing. This is a step towards understanding the host mass dependence of the exoplanet mass ratio function.

Creating a large sample of low host-star masses and their companions is of crucial importance for occurrence rate measurement studies \citep{Pass_2023} and for population synthesis models, as it provides a more complete exploration of the parameter space of the observational detection bias used \citep{Emsenhuber2023EPJP..138..181E}. The high sensitivity of gravitational microlensing to detect companion planets to this type of stars, in combination to high angular resolution follow-up observations promises a large number of high precision planet detections with Nancy Grace Roman Space Telescope.

Another notable point we infer from this work is that OGLE-2013-BLG-0132 is a perfect candidate for an HST follow-up observations. The results of the different observations and analysis methods of this study are in absolute agreement between them with very high precision. This makes this event an excellent candidate to test and ameliorate our techniques of measuring the planet and host star masses with/for Nancy Grace Roman Space Telescope. Our Keck images show a clear separation between source and lens, observations in different bands with Hubble Space Telescope will help us better acquire the systematic error sources in our methods. Finally, measuring the microlensing parallax for short length events can be difficult, even by processing different observational bands, which makes this target even more interesting for testing and validating the parallax measurement methods.
\clearpage

\begin{acknowledgments}
{N.E.R. would like to acknowledge Mr. Ioannis Vartholomeos for his precious help during this work, by providing interesting discussions and fruitful questions about the gravitational microlensing technique. Unfortunately, Mr. Vartholomeos passed away this summer and didn't have the chance to see this work completed.

This work was supported by the University of Tasmania through the UTAS Foundation and the endowed Warren Chair in Astronomy and the ANR COLD-WORLDS (ANR-18-CE31-0002) and by NASA through grant NASA-80NSSC18K0274. This research was also supported by the Australian Government through the Australian Research Council Discovery Program (project number 200101909) grant awarded to AC and JPB. 
The Keck Telescope observations and analysis were supported by a NASA Keck PI Data Award, administered by the NASA Exoplanet Science Institute. Data presented herein were obtained at the W. M. Keck Observatory from telescope time allocated to the National Aeronautics and Space Administration through the agency’s scientific partnership with the California Institute of Technology and the University of California. The Observatory was made possible by the generous financial support of the W. M. Keck Foundation. The authors wish to recognize and acknowledge the very significant cultural role and reverence that the summit of Mauna Kea has always had within the indigenous Hawaiian community. We are most fortunate to have the opportunity to conduct observations from this mountain.}
\end{acknowledgments}

\bibliography{OB132}{}
\bibliographystyle{aasjournal}
\end{document}